# State-Aware IoT Scheduling Using Deep Q-Networks and Edge-Based Coordination

Qingyuan He
New York University
New York, USA

Chang Liu
Washington University in St. Louis
St. Louis, USA

Juecen Zhan
Vanderbilt University
Nashville, USA

Weiqiang Huang
Northeastern University
Boston, USA

Ran Hao*
University of North Carolina at Chapel Hill
Chapel Hill, USA

*Abstract-This paper addresses the challenge of energy efficiency management faced by intelligent IoT devices in complex application environments. A novel optimization method is proposed, combining Deep Q-Network (DQN) with an edge collaboration mechanism. The method builds a state-action-reward interaction model and introduces edge nodes as intermediaries for state aggregation and policy scheduling. This enables dynamic resource coordination and task allocation among multiple devices. During the modeling process, device status, task load, and network resources are jointly incorporated into the state space. The DQN is used to approximate and learn the optimal scheduling strategy. To enhance the model's ability to perceive inter-device relationships, a collaborative graph structure is introduced to model the multi-device environment and assist in decision optimization. Experiments are conducted using real-world IoT data collected from the FastBee platform. Several comparative and validation tests are performed, including energy efficiency comparisons across different scheduling strategies, robustness analysis under varying task loads, and evaluation of state dimension impacts on policy convergence speed. The results show that the proposed method outperforms existing baseline approaches in terms of average energy consumption, processing latency, and resource utilization. This confirms its effectiveness and practicality in intelligent IoT scenarios.*

*Keywords-Deep reinforcement learning, edge computing, energy efficiency optimization, IoT scheduling.*

## I. INTRODUCTION

With the rapid development of the Internet of Things (IoT), a growing number of smart devices have been widely deployed in various scenarios, including urban management, industrial manufacturing, home security, and smart healthcare. These devices play a crucial role in enhancing service quality, optimizing resource allocation, and improving system intelligence. However, the large-scale deployment of IoT devices has also led to a significant increase in energy consumption [1]. According to relevant statistics, the energy consumption of global IoT terminal devices accounts for a substantial portion of the total energy used by communication networks. Reducing energy consumption while maintaining system performance has become a key challenge for the sustainable development of intelligent IoT systems. The traditional centralized cloud computing model cannot meet the dual demands of low latency and high energy efficiency. A new collaborative mechanism is urgently needed to balance computing resource allocation and energy control.

Edge computing, as a data processing paradigm close to end devices, provides a new approach to energy efficiency optimization in intelligent IoT systems. By migrating computing resources from remote data centers to edge nodes near data sources, edge computing effectively reduces communication latency and enables partial local processing. This significantly lowers the energy cost caused by frequent remote interactions. More importantly, the deployment of edge nodes facilitates information sharing and resource collaboration among multiple devices [2]. This makes cross-device load balancing and joint scheduling feasible. However, due to the dynamic nature of device states, varying task requirements, and heterogeneous network conditions, energy efficiency optimization in edge environments is highly dynamic and uncertain. Traditional static scheduling methods struggle to handle the complexity of real-world scenarios.

Against this background, Deep Reinforcement Learning (DRL) has emerged as a promising solution for dynamic energy efficiency optimization. DRL is known for its adaptability, online learning capabilities [3], and policy optimization [4]. In particular, Deep Q-Networks (DQN) leverage deep neural networks to approximate the state-action value function and have been applied in Human-Computer Interaction for adaptive interfaces [5-6], in NLP for entity extraction [7], and in LLM systems for optimizing multi-turn interaction strategies [8-9]. This enables the discovery of optimal strategies in high-dimensional state spaces and supports dynamic energy management for smart devices in diverse environments. Compared to traditional optimization algorithms, DQN demonstrates superior convergence and robustness when facing incomplete system states, delayed feedback signals, and complex policy combinations. It also has strong generalization abilities and practical deployment value.

However, most existing studies applying DQN to energy efficiency optimization in IoT focus on single-device or local strategies. These approaches often ignore the collaborative relationships among devices, limiting the overall system potential. Therefore, developing an integrated optimization framework that combines DQN with edge collaboration mechanisms can improve global energy efficiency management. It can also enable resource sharing, load awareness, and complementary strategies across devices. This would help reduce total system energy consumption and improve resource utilization efficiency. Moreover, a DQN framework with edge collaboration is adaptable to heterogeneous network environments. It supports asynchronous coordination among devices and multi-objective optimization, making it a promising core scheduling strategy for future intelligent IoT systems.

This study focuses on smart IoT devices and explores the deep integration of DQN-based reinforcement learning with edge collaboration mechanisms. The goal is to propose a joint scheduling method with adaptive learning, resource coordination, and energy optimization capabilities. This research is not only practically significant for improving the operational efficiency of existing IoT systems but also provides theoretical support for the application of reinforcement learning in energy management under distributed and heterogeneous environments. By optimizing the DQN structure and designing a collaborative edge strategy, this study aims to promote the development of intelligent IoT systems toward efficiency, sustainability, and intelligence. It provides technological support for building a resource-saving and environmentally friendly information society.

## II. METHOD

This study built an energy efficiency optimization model based on DQN and integrated the edge collaboration mechanism to achieve joint scheduling and energy consumption management among multiple devices. The model architecture is shown in Figure 1.

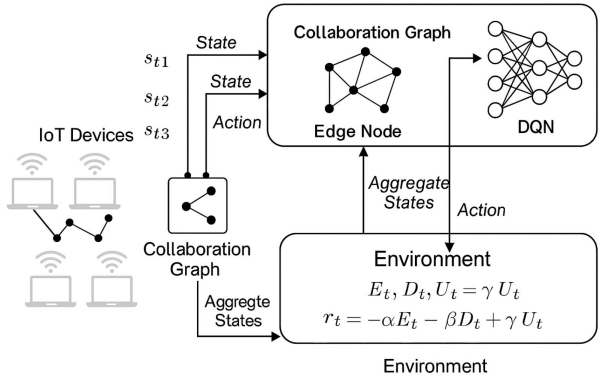

Figure 1. Overall model architecture

Figure 1 shows the overall architecture of the energy efficiency optimization model based on DQN and edge collaboration mechanism proposed in this study. IoT devices aggregate local state information to edge nodes through collaborative graph structures. Edge nodes build collaborative relationships between devices and form joint state vectors that are input into the DQN network for action decision-making. Ultimately, the agent performs actions in the environment and generates reward feedback based on indicators such as energy consumption, latency, and resource utilization, driving continuous optimization of strategies.

In system modeling, each smart IoT device is first modeled as an agent, whose state is represented by multi-dimensional features such as the current energy consumption level of the device, the remaining battery power, the length of the task queue, and the resource occupancy of the edge node, forming a state space S. The action space A is defined as the task processing method selected by the device at each time step, including options such as "local processing", "edge offloading", and "delayed processing". After each state-action interaction of the system, the immediate reward $r_t$ obtained is defined as a comprehensive indicator related to energy efficiency, and its expression is:

$$r_t = -\alpha E_t - \beta D_t + \gamma U_t$$

$E_t$ represents the current energy consumption, $D_t$ represents the processing delay, $U_t$ represents the edge resource utilization, and $\alpha$、$\beta$、$\gamma$ is the experience weight coefficient. This reward function is designed to encourage the agent to improve resource utilization efficiency while reducing energy consumption and delay, reflecting the comprehensive goal of energy efficiency optimization.

Building upon the previously established model, this study employs a Deep Q-Network (DQN) to approximate the optimal policy function for dynamic resource coordination in the multi-device IoT environment. Inspired by the reinforcement learning framework outlined by Sun [10], the agent continuously interacts with the environment to capture state transitions reflecting task load, device status, and network resource changes. To enhance learning stability and efficiency, the experience replay mechanism is adopted, which decouples data correlation and promotes robust convergence—an approach that aligns with the sampling strategies recommended by Wang [11] in time series modeling tasks. The Q-value function is refined through iterative updates governed by the Bellman optimality principle, enabling progressive policy improvement. This optimization step is further supported by neural approximation techniques similar to those discussed in Li [12], where model generalization across dynamic input distributions is achieved through efficient representation learning. The resulting update rule takes the following form:

$$Q(s_t, a_t) = r_t + \gamma \max_{a'} Q(s_{t+1}, a')$$

Where $\gamma$ is the discount factor, which controls the impact of long-term returns. DQN uses a multi-layer perceptron (MLP) to model $Q(s,a)$ and minimizes the mean squared error loss function:

$$L(\theta) = E_{(s,a,r,s')}[(y_t - Q(s,a;\theta))^2]$$

$$y_t = r + \gamma \max_{a'} Q(s', a', \theta^-)$$

To update the network parameters $\theta$, where $\theta^-$ represents the target network parameters, it is regularly synchronized with the main network to improve training stability.

To further enhance the system's collaborative capabilities in a multi-device IoT environment, this study introduces an edge collaborative mechanism designed to enable real-time information sharing and coordinated policy execution. Drawing on the work of Wang [13], who emphasizes adaptive weighting strategies in distributed environments, the proposed mechanism leverages edge nodes to aggregate local device states and distribute optimized scheduling policies. This decentralized structure supports timely decision-making under fluctuating network and task conditions. In line with the representational techniques highlighted by Guo et al. [14], the system architecture incorporates modular components that facilitate self-supervised adaptation and local inference, enabling effective collaboration without central dependency. Additionally, inspired by the communication optimization principles outlined by Sun [15], the mechanism supports dynamic interfacing across heterogeneous devices through lightweight coordination protocols. Specifically, edge nodes periodically summarize the status information of surrounding devices, and model the spatial adjacency relationship between devices through a graph structure to form a collaborative graph $G(V, E)$. Edge nodes assist local devices in optimizing their action decisions by aggregating the status information of adjacent devices, thereby forming a strategy close to the global optimal from a local perspective. Under this mechanism, the agent's Q function is expanded to a conditional Q function $Q(s_t^i, a_t^i | N_t^i)$, where $s_t^i$ and $a_t^i$ are the state and action of the i-th device, respectively, and $N_t^i$ represents the set of its neighbor device states. This modeling method not only improves the context-awareness of Q-value estimation, but also effectively enhances the coordination of strategies and system-level energy efficiency performance.

## III. EXPERIMENT

### A. Datasets

The dataset used in this study is sourced from the FastBee intelligent device access and management platform. This platform integrates a large number of real-world deployed IoT terminal nodes, including environmental monitoring devices, video capture modules, smart plugs, and low-power sensors. These devices communicate in real time with edge nodes through a unified access protocol. Data are collected every 30 seconds, covering multiple dimensions such as device operation status, task request frequency, battery level changes, resource usage ratio, and edge node response feedback. These raw data provide complete support for constructing device state spaces, action decision scenarios, and energy feedback mechanisms.

To meet the modeling requirements of the state-action-reward triplet in reinforcement learning, the raw data were preprocessed. First, operational features from different devices were normalized to reduce the impact of scale differences. Then, device state sequences were constructed based on time windows, and the energy consumption and latency after each action were labeled as the basis for immediate rewards. State variables include current device load, remaining battery level, task queue length, and network condition indicators. Action dimensions involve decisions such as local processing, edge offloading, and task delay. This dataset reflects the energy performance of devices under current operating conditions.

In addition, to simulate a realistic multi-device joint scheduling scenario in a collaborative environment, this study incorporated the states of multiple edge nodes and their communication topology. A collaboration graph was built to define adjacency relationships among devices and their resource interaction patterns. This further enhances the dataset's suitability for modeling edge collaboration strategies. The final dataset captures both the dynamic evolution of individual device states and the energy efficiency behavior sequences under multi-device collaboration. It provides realistic, rich, and representative data support for training and testing DQN models in complex environments.

### B. Experimental Results

This paper first gives an energy consumption comparison experiment under different scheduling strategies, and the experimental results are shown in Table 1.

Table 1. Energy consumption comparison experiment under different scheduling strategies

| Scheduling strategy | Average energy consumption (mWh) | Average processing delay (ms) | Edge resource utilization (%) |
|---|---|---|---|
| Local independent processing | 58.42 | 83.7 | 12.4 |
| All Edge | 39.85 | 64.1 | 86.7 |
| Random Scheduling Strategy[16] | 47.63 | 91.5 | 49.3 |
| Static round-robin scheduling[17] | 44.27 | 78.9 | 65.8 |
| DQN + Edge | 31.96 | 55.3 | 88.5 |

The experimental results show significant differences in performance across various scheduling strategies in terms of energy consumption, processing latency, and resource utilization. The traditional "local-only processing" strategy lacks task offloading capability. As a result, each device must complete all computation tasks independently. This leads to the highest average energy consumption, reaching 58.42 mWh. It also suffers from high processing delay and extremely low edge resource utilization, only 12.4%. In contrast, the "full edge offloading" strategy achieves improvements in energy consumption and latency. However, it relies heavily on edge nodes, causing resource saturation. It lacks flexible scheduling and shows limited system robustness.

Random scheduling and static polling strategies moderately improve resource utilization. However, their overall energy efficiency remains limited. In particular, the random strategy shows high uncertainty in scheduling due to the absence of a learning mechanism. Its average energy consumption remains relatively high at 47.63 mWh, and it results in the worst delay performance. The static polling strategy follows a fixed task assignment logic and shows slight improvement in energy efficiency. Yet, it cannot perceive device states in real time, making it difficult to adapt to dynamic workloads. The system's overall performance remains suboptimal.

In contrast, the proposed "DQN + edge collaboration" strategy outperforms all others across all three dimensions. Driven by the reinforcement learning mechanism, the agent dynamically adjusts scheduling policies based on past interaction experiences. It enables precise task offloading and resource coordination. The introduction of the collaboration graph further enhances information sharing and policy complementarity among devices. As a result, the system's average energy consumption drops to 31.96 mWh, processing latency is reduced to 55.3 ms, and edge resource utilization reaches 88.5%. These results demonstrate the effectiveness and practicality of the proposed method for energy efficiency optimization in multi-device environments.

Furthermore, this paper presents a model robustness verification experiment under heterogeneous task loads, and the experimental results are shown in Figure 2.

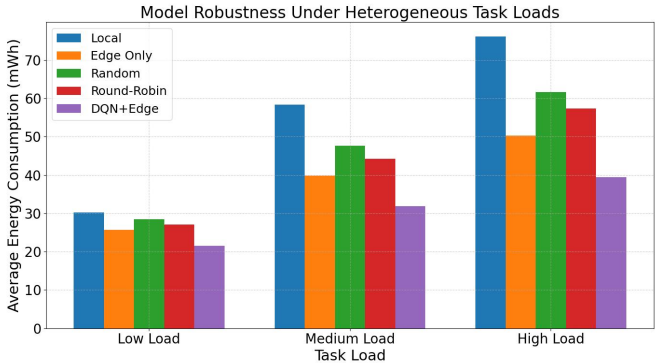

Figure 2. Model robustness under heterogeneous task loads

As shown in Figure 2, the energy consumption of all scheduling strategies increases as task load rises from low to high. However, their sensitivity to load variation differs significantly. The "Local" strategy consistently shows high energy consumption under all three load levels. In high-load scenarios, its average consumption approaches 75 mWh. This reflects its lack of flexible scheduling ability, low resource utilization efficiency, and poor system robustness.

In contrast, the "Edge Only," "Random," and "Round-Robin" strategies perform better in low and medium loads. But as the load increases, their ability to control energy efficiency declines. The "Random" strategy, in particular, shows large fluctuations under high load. This indicates instability and weak adaptability in its scheduling decisions. Although "Round-Robin" is more stable overall, it cannot dynamically adjust based on system states and thus fails to achieve optimal energy efficiency.

The "DQN + Edge" strategy maintains the best energy performance under all task loads. It demonstrates strong robustness. Even under high load, its energy consumption remains significantly lower than other strategies. This shows that the method has strong adaptability and flexible scheduling capabilities. The DQN reinforcement learning framework, combined with the edge collaboration structure, enables the model to select optimal actions according to environmental changes. This balances energy efficiency and system performance, verifying the practical value of the method in heterogeneous task environments.

Finally, this paper presents an experiment on the impact of state space dimension on strategy convergence speed, and the experimental results are shown in Figure 3.

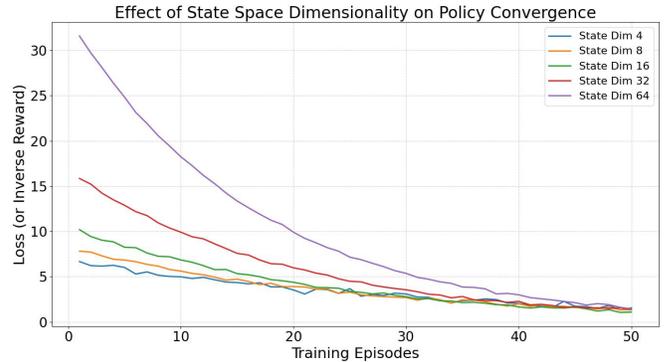

Figure 3. The impact of state space dimension on strategy convergence speed

As shown in Figure 3, the convergence speed of the policy decreases as the dimensionality of the state space increases. This trend is especially noticeable in the early stages of training. Models with state dimensions of 4 and 8 show rapid loss reduction within the first 20 training episodes. Their convergence is fast, and the loss curves are smooth and steep. In contrast, the model with a state dimension of 64 starts with a higher initial loss and shows a much slower decline. This indicates that a high-dimensional state space significantly increases the complexity of policy learning and affects training efficiency.

Although all models eventually converge after sufficient training, high-dimensional state spaces introduce redundant information and noise. This slows down convergence and may also affect model stability and generalization. In comparison, medium-dimensional settings (such as 16 and 32) achieve a better balance between training stability and policy representation. They demonstrate more favorable overall performance. These findings confirm that the choice of state space dimensionality plays a critical role in both policy effectiveness and convergence efficiency.

IV. CONCLUSION

This paper addresses the problem of energy efficiency optimization in intelligent IoT environments. A scheduling method combining Deep Q-Network (DQN) and edge collaboration mechanisms is proposed. By enabling dynamic state awareness and joint decision-making across multiple devices, the method effectively improves energy usage efficiency and task processing performance. It integrates the adaptability of reinforcement learning with the collaborative advantages of edge computing. This addresses the limitations of traditional scheduling strategies, which often suffer from high energy consumption and poor adaptability in complex and heterogeneous device environments.

In terms of method design, an optimization framework based on the state-action-reward structure is constructed. A collaborative graph is introduced to model interactions between devices, allowing the DQN to make more accurate action selections under multi-source state inputs. Through extensive simulation experiments, the proposed method demonstrates superior performance in energy control, task delay, and resource utilization. It shows strong robustness and convergence ability, especially under multi-load and heterogeneous conditions, validating its potential for real-world IoT applications. Despite the promising results, several areas remain for further exploration. The current reward function is still based on a linear combination and does not incorporate multi-objective trade-offs. Additionally, the scalability of DQN in high-dimensional state spaces is limited. Future work could explore the use of distributed multi-agent reinforcement learning [18], graph neural networks [19], or policy gradient methods [20] to enhance system adaptability and generalization. It is also important to validate the model in larger-scale, multi-node edge computing environments. Future research can further focus on stable scheduling in dynamic network conditions for intelligent IoT systems. The integration of energy harvesting, task prioritization, and adaptive offloading mechanisms should be explored. In addition, combining federated learning can enable privacy-preserving model training, meeting the dual demands of computational efficiency and data privacy at the edge. These efforts will support the practical deployment of intelligent scheduling algorithms on a wider scale.

REFERENCES

[1] X. Kong, et al., "Deep reinforcement learning-based energy-efficient edge computing for internet of vehicles," IEEE Transactions on Industrial Informatics, vol. 18, no. 9, pp. 6308–6316, 2022.

[2] T. Zheng, et al., "Deep reinforcement learning-based workload scheduling for edge computing," Journal of Cloud Computing, vol. 11, no. 1, pp. 3, 2022.

[3] Y. Deng, "A Reinforcement Learning Approach to Traffic Scheduling in Complex Data Center Topologies," Journal of Computer Technology and Software, vol. 4, no. 3, 2025.

[4] Y. Wang, "Optimizing Distributed Computing Resources with Federated Learning: Task Scheduling and Communication Efficiency," Journal of Computer Technology and Software, vol. 4, no. 3, 2025.

[5] Q. Sun, "Dynamic Optimization of Human-Computer Interaction Interfaces Using Graph Convolutional Networks and Q-Learning," Transactions on Computational and Scientific Methods, vol. 5, no. 2, 2025.

[6] S. Duan, "Deep Learning-Based Gesture Key Point Detection for Human-Computer Interaction Applications," Transactions on Computational and Scientific Methods, vol. 5, no. 1, 2025.

[7] X. Wang, G. Liu, B. Zhu, J. He, H. Zheng and H. Zhang, "Pre-trained Language Models and Few-shot Learning for Medical Entity Extraction," arXiv preprint arXiv:2504.04385, 2025.

[8] A. Kai, L. Zhu and J. Gong, "Efficient Compression of Large Language Models with Distillation and Fine-Tuning," Journal of Computer Science and Software Applications, vol. 3, no. 4, pp. 30–38, 2023.

[9] G. Cai, J. Gong, J. Du, H. Liu and A. Kai, "Investigating Hierarchical Term Relationships in Large Language Models," Journal of Computer Science and Software Applications, vol. 5, no. 4, 2025.

[10] X. Sun, "Dynamic Distributed Scheduling for Data Stream Computing: Balancing Task Delay and Load Efficiency," Journal of Computer Technology and Software, vol. 4, no. 1, 2025.

[11] J. Wang, "Multivariate Time Series Forecasting and Classification via GNN and Transformer Models," Journal of Computer Technology and Software, vol. 3, no. 9, 2024.

[12] P. Li, "Improved Transformer for Cross-Domain Knowledge Extraction with Feature Alignment," Journal of Computer Science and Software Applications, vol. 5, no. 2, 2024.

[13] X. Wang, "Mining Multimodal Data with Sparse Decomposition and Adaptive Weighting," Transactions on Computational and Scientific Methods, vol. 5, no. 1, 2025.

[14] F. Guo, X. Wu, L. Zhang, H. Liu and A. Kai, "A Self-Supervised Vision Transformer Approach for Dermatological Image Analysis," Journal of Computer Science and Software Applications, vol. 5, no. 4, 2025.

[15] Q. Sun, "A Visual Communication Optimization Method for Human-Computer Interaction Interfaces Using Fuzzy Logic and Wavelet Transform," Proceedings of the 2024 4th International Conference on Communication Technology and Information Technology (ICCTIT), pp. 140–144, Dec. 2024.

[16] H. Hua, et al., "Edge computing with artificial intelligence: A machine learning perspective," ACM Computing Surveys, vol. 55, no. 9, pp. 1–35, 2023.

[17] Z. Li, et al., "Sparse trace ratio LDA for supervised feature selection," IEEE Transactions on Cybernetics, vol. 54, no. 4, pp. 2420–2433, 2023.

[18] Z. Liu, J. Zhang, E. Shi, Z. Liu, D. Niyato, B. Ai and X. S. Shen, "Graph neural network meets multi-agent reinforcement learning: Fundamentals, applications, and future directions," IEEE Wireless Communications, 2024.

[19] A. Liang, "A Graph Attention-Based Recommendation Framework for Sparse User-Item Interactions," Journal of Computer Science and Software Applications, vol. 5, no. 4, 2025.

[20] E. Hosseini, L. Reinhardt and D. B. Rawat, "Optimizing gradient methods for IoT applications," IEEE Internet of Things Journal, vol. 9, no. 15, pp. 13694–13704, 2022.